\documentclass[lettersize,journal]{IEEEtran}
\IEEEoverridecommandlockouts
\usepackage{cite}
\usepackage{amsmath,amssymb,amsfonts}
\usepackage{amsthm}

\usepackage{bm}
\usepackage{graphicx}
\usepackage{textcomp}
\usepackage{algorithm}
\usepackage{algorithmic}
\usepackage{xcolor}
\usepackage[caption=false,font=normalsize,labelfont=sf,textfont=sf]{subfig}
\usepackage{booktabs}
\usepackage{multirow}
\usepackage{makecell}
\usepackage{array}
\usepackage{stfloats}
\usepackage{url}
\usepackage{verbatim}
\AtBeginDocument{\hfuzz=1000pt}

\def\BibTeX{{\rm B\kern-.05em{\sc i\kern-.025em b}\kern-.08em
T\kern-.1667em\lower.7ex\hbox{E}\kern-.125emX}}
\begin{document}

%\title{CR$^2$: Cost-Aware Risk-Controlled Routing for Two-Stage LLM Inference at the Mobile Edge}
\title{CR$^2$: Cost-Aware Risk-Controlled Routing for Wireless Device-Edge LLM Inference}
\author{Nan Xue, Shengkang Chen, Zhiyong Chen,~\IEEEmembership{Senior Member,~IEEE,} Jiangchao Yao, Yaping Sun, \\ Zixia Hu, and Meixia Tao, \IEEEmembership{Fellow,~IEEE}
\thanks{N. Xue, S. Chen, Z. Chen, J. Yao and M. Tao are with the Cooperative Medianet Innovation
			Center, Shanghai Jiao Tong University, Shanghai 200240, China. (email:
			\{nan.xue, chen20, zhiyongchen, Sunarker, mxtao\}@sjtu.edu.cn).}%
\thanks{Y. Sun is with the Department of Broadband Communication, Pengcheng
			Laboratory, Shenzhen 518000, China. Y. Sun is also
			with the Future Network of Intelligent Institute (FNii), the Chinese University
			of Hong Kong (Shenzhen), Shenzhen 518172, China (email: sunyp@pcl.ac.cn).}%
\thanks{Z. Hu is with Meta, Menlo Park, CA 94025 USA (email: huzixia@meta.com).}%
}
\maketitle

\begin{abstract}
As large language models (LLMs) move from centralized clouds to mobile edge environments, efficient serving must balance latency, energy consumption, and accuracy under constrained device–edge resources. Query-level routing between lightweight on-device models and stronger edge models provides a flexible mechanism to navigate this trade-off. However, existing routers are designed for centralized cloud settings and optimize token-level costs, failing to capture the dynamic latency and energy overheads in wireless edge deployments. In this paper, we formulate mobile edge LLM routing as a deployment-constrained, cost-aware decision problem, and propose CR$^2$, a two-stage device–edge routing framework. CR$^2$ decouples a lightweight on-device margin gate from an edge-side utility selector for deferred queries. The margin gate operates on frozen query embeddings and a user-specified cost weight to predict whether local execution is utility-optimal relative to the best edge alternative under the target operating point. We further introduce a conformal risk control (CRC) calibration procedure that maps each operating point to an acceptance threshold, enabling explicit control of the marginal false-acceptance risk under the full-information utility reference. Experiments on the routing task show that CR$^2$ closely matches a full-information reference router using only device-side signals before deferral. Compared with strong query-level baselines, CR$^2$ consistently improves the deployable accuracy--cost Pareto frontier and reduces normalized deployment cost by up to 16.9\% at matched accuracy.
\end{abstract}


\bstctlcite{IEEEexample:BSTcontrol}
\section{Introduction}
\IEEEPARstart{L}arge language models (LLMs) have achieved strong performance on language understanding, reasoning, and generation~\cite{gpt4}, and are increasingly moving from centralized clouds to mobile-edge platforms~\cite{qu2025mobile}. This shift is driven by latency, privacy, and service-continuity requirements that favor inference close to users. Meanwhile, lightweight model families such as Qwen~\cite{qwen3}, together with mature on-device inference stacks~\cite{ollama2026}, enable LLM execution on both user equipment (UE) and edge servers. These trends create a new design space for mobile-edge collaborative inference, where routing decisions must account for heterogeneous device capabilities, time-varying wireless conditions, strict latency–energy constraints and limited decision-time information. Importantly, in the device--edge setting, these constraints induce a \emph{two-stage decision structure}, where local execution decisions must be made at the UE using only locally available information before deferring queries to the edge.

Prior work on mobile-edge LLM systems has explored caching, training, and inference across UE and edge servers~\cite{qu2025mobile}. Systems such as EdgeShard~\cite{zhang2024edgeshard} and CE-CoLLM~\cite{jin2024cecollm} show that deployment decisions are shaped by model partitioning, contextual data transfer, and hardware heterogeneity. Long-context edge serving further couples inference decisions with cache state and context evolution~\cite{xu2025longcontextedge}. These studies indicate that mobile-edge LLM serving is fundamentally a communication–computation co-design problem.

A related line of work investigates collaborative inference mechanisms such as split inference and speculative decoding. While effective in reducing computation on individual devices, these methods require repeated coordination, making end-to-end latency highly sensitive to bandwidth fluctuations and round-trip delays~\cite{kang2017neurosurgeon,laskaridis2020spinn,li2018edgent,leviathan2023fast,chen2024sequoia,li2024eagle}. This motivates query-level routing mechanisms that make a one-shot decision and avoid dense cross-tier coordination.

In this context, query-level routing over complete models is attractive because it selects a single execution path per request and avoids inter-tier synchronization. However, existing routing methods are primarily designed for centralized model pools. They optimize static quality-cost trade-offs and rely on token-level cost proxies, without modeling deployment-induced communication and latency constraints. Meanwhile, existing edge LLM systems focus on runtime resource coordination but do not explicitly address the two-stage routing structure, where local decisions must be made before edge-side utility estimates and runtime edge-state information are available.

The two-stage structure introduces asymmetric decision risk. Unnecessary deferral incurs additional latency and energy consumption due to communication overhead, whereas false local acceptance irreversibly forgoes edge models and may degrade user experience. Therefore, the local decision rule should be conservative, estimate the relative utility of local execution under a given cost-accuracy trade-off, and provide a controllable mechanism to regulate the marginal risk of false local acceptance.

To address this gap, we propose CR$^2$, a device-edge collaborative routing framework for device-edge LLM inference. CR$^2$ decomposes routing into two stages. At the UE, a lightweight \emph{margin gate} computes a utility margin that estimates whether local execution is competitive with the best edge alternative under a user-specified cost weight. Accepted queries are executed locally, while deferred queries are forwarded to the edge server, where a utility-based selector chooses among available models. To ensure reliable deployment, we introduce a conformal risk control (CRC) calibration mechanism~\cite{angelopoulos2024conformal}, which maps each operating point to an acceptance threshold, enabling explicit control of the marginal false-acceptance risk under the routing utility. The UE module requires only a frozen encoder and a lightweight scoring head, incurring minimal overhead before local inference.

The main contributions of this paper are summarized as follows:
\begin{itemize}
\item We establish a deployment-constrained two-stage routing framework for device-edge LLM inference. The framework separates UE local acceptance from edge model selection under state-dependent latency and energy costs, while respecting the constraint that the device cannot access edge model scores before deferral.

\item We develop CR$^2$, which realizes this framework through a UE-observable margin gate that distills full-information utility margins into a deployable on-device score, a monotonicity regularizer that promotes coherent decisions across cost-weight operating points, and a per-$\lambda$ CRC calibration rule that controls the marginal false-acceptance risk under the fixed learned score.

\item We evaluate CR$^2$ on a benchmark-derived routing task with profiled UE and edge server deployment cost models. Compared with KNN, MLP, EmbedLLM, and LLMRank, CR$^2$ achieves the best deployable Pareto frontier and reduces normalized deployment cost by up to $16.9\%$ at matched accuracy.
\end{itemize}

The rest of this paper is organized as follows. Sec.~II reviews related work, and Sec.~III summarizes the CRC calibration primitive. Sec.~IV presents the mobile-edge system and deployment-cost model. Sec.~V formulates and analyzes the deployment-constrained routing framework. Sec.~VI develops the CR$^2$ method, Sec.~VII reports the experimental results, and Sec.~VIII concludes the paper.

\section{Related Work}

\subsubsection{Query-level LLM routing}
Dynamic routing for LLM efficiency spans multiple granularities, including token-level mixtures-of-experts~\cite{fedus2022switch,zhou2022mixture,li2025rt} and generation-time mechanisms such as speculative decoding~\cite{leviathan2023fast,lu2311routing,chen2024sequoia,li2024eagle,jiang2026trispec}. Our work focuses on query-level routing, which assigns each request to a single model from a candidate pool. Representative methods learn lightweight routing policies from preference data, correctness labels, or system-level cost signals~\cite{ong2025routellm,ding2024hybrid,feng2025graphrouter,stripelis2024tensoropera,ding2025best,bao2025dynamicrouting}. Other approaches reduce supervision requirements via label-free routing~\cite{guha2024smoothie} or capability-aware training~\cite{zhang2025capability}. Cascade-based methods such as FrugalGPT~\cite{chen2024frugalgpt} instead escalate across models until a quality threshold is satisfied. Edge-specific routing has also optimized QoS-aware assignment among multiple edge LLM experts under dynamic workload states~\cite{yang2025qosrouting}.

\subsubsection{Single-model inference at the edge}
A parallel line of work studies single-model inference between mobile devices and edge servers to meet latency and energy constraints. Neurosurgeon~\cite{kang2017neurosurgeon} and SPINN~\cite{laskaridis2020spinn} partition a single CNN at run time; Edgent~\cite{li2018edgent} performs adaptive DNN partitioning for 5G environments. EdgeBERT~\cite{tambe2021edgebert} co-designs a transformer with early-exit gates for mobile inference. These systems optimize the execution of one model under a fixed computation graph, whereas the proposed CR$^2$ selects among heterogeneous complete LLMs under a two-stage deployment structure.

\subsubsection{Edge LLM serving and collaboration}
Several recent works study LLM-specific serving and deployment at the edge. MEI4LLM surveys edge caching, training, and inference for LLM services~\cite{qu2025mobile}. EdgeShard and CE-CoLLM study collaborative LLM execution under heterogeneous edge resources and contextual-data transfer costs~\cite{zhang2024edgeshard,jin2024cecollm}. Other studies consider long-context model caching and inference offloading, MoE expert caching, wireless distributed MoE deployment and training, cold-start scheduling, wireless-informed distributed speculative decoding, or collaborative inference over heterogeneous edge LLMs under latency and energy constraints~\cite{xu2025longcontextedge,chen2026slimcaching,xue2024wdmoe,shi2025stablemoe,liu2025csgo,liu2026wisv,jin2025moe2}. Wireless AI work also uses LLMs to assist intent-driven resource allocation~\cite{sun2026llmempowered}, while our setting treats LLM inference as the workload to be routed.

\subsubsection{Representation learning for routing}
Several recent routers use learned embeddings to anticipate which model can solve a query. EmbedLLM~\cite{zhuang2025embedllm} trains a textual matrix factorization that jointly embeds queries and model accuracy profiles; RouterDC~\cite{chen2024routerdc} and RadialRouter~\cite{jin2025radialrouter} use contrastive objectives over query--model pairs. The proposed CR$^2$ keeps the query encoder frozen and learns a continuous utility-margin surrogate for UE execution relative to the best edge alternative, conditioned jointly on the query embedding and the operator cost weight $\lambda$. The resulting gate is evaluated solely for the local acceptance decision and never requires edge model utility estimates before deferral.

\subsubsection{Risk-controlled decision systems}
Conformal prediction~\cite{vovk2005algorithmic} and its risk-control extensions~\cite{angelopoulos2022learn,angelopoulos2024conformal} provide distribution-free tools for calibrating predictive systems under coverage or bounded-risk criteria. These tools are typically used to calibrate the output of an already specified predictor or prediction set. In contrast, CR$^2$ uses CRC to calibrate a local acceptance gate whose accepted set is induced under partial information, before edge model scores are available.

In summary, query-level LLM routers and edge LLM serving systems address complementary parts of this problem. The former select among complete models but usually assume a centralized candidate model pool, token-level cost proxies, or an edge-side router with global QoS state; the latter model deployment factors such as hardware heterogeneity, communication overhead, and latency--energy trade-offs, but focus mainly on execution and resource coordination. CR$^2$ connects these lines by formulating query-level model selection under a two-stage device-edge deployment: the UE accepts locally without edge-model scores, and only deferred queries are routed to the edge model pool.

\section{Preliminaries}
\label{sec:preliminaries}

This section reviews CRC~\cite{angelopoulos2024conformal}, which we use to calibrate the UE-side acceptance decision in the proposed two-stage routing framework.

Let $\mathcal{D}_{\mathrm{cal}}=\{(x_i,y_i)\}_{i=1}^{n}$ be a calibration set, and let $(x_{n+1},y_{n+1})$ denote a future test sample, where $\{(x_i,y_i)\}_{i=1}^{n+1}$ are exchangeable. CRC considers a family of post-processing rules $\{C_{\eta}\}_{\eta\in\Lambda}$ parameterized by $\eta$, together with a bounded loss function $\ell(C_\eta(x),y)\le B$ that is non-increasing in $\eta$.

Given a target risk level $\alpha$, CRC selects
\begin{equation}
\hat{\eta}
=
\inf\left\{
\eta\in\Lambda:
\frac{n}{n+1}\widehat{R}_n(\eta)+\frac{B}{n+1}\le \alpha
\right\},
\label{eq:crc_eta}
\end{equation}
where
\begin{equation}
\widehat{R}_n(\eta)
=
\frac{1}{n}\sum_{i=1}^{n}
\ell(C_\eta(x_i),y_i)
\end{equation}
is the empirical risk on the calibration set. Under exchangeability, boundedness, and monotonicity, the deployed prediction object $C_{\hat\eta}$ satisfies
\begin{equation}
\mathbb{E}\!\left[
\ell\!\left(C_{\hat\eta}(x_{n+1}),y_{n+1}\right)
\right]\le \alpha.
\label{eq:crc_guarantee}
\end{equation}

CRC thus provides a distribution-free mechanism to control expected decision risk under minimal assumptions.

In this work, CRC is used to calibrate the UE-side acceptance threshold in the two-stage routing process. The task-specific loss function and threshold-selection rule are introduced in Sec.~\ref{sec:crc}.

\section{System Model}
\label{sec:sysmodel}

\begin{figure}[t]
  \centering
  \includegraphics[width=0.48\textwidth]{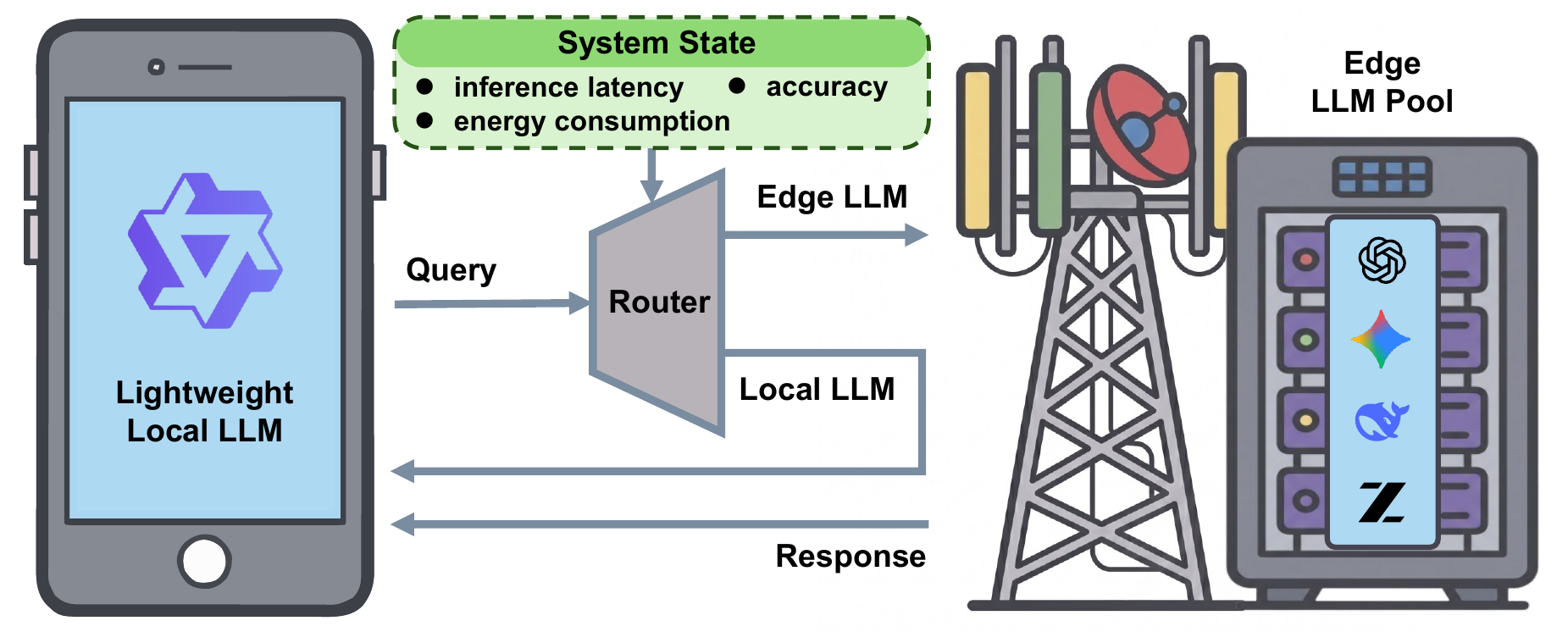}
  \caption{Two-tier device-edge inference and routing flow.}
  \label{fig:framework}
\end{figure}

\subsection{Deployment Architecture and Information Structure}

We consider a two-tier collaborative inference system consisting of a UE and an edge server, as illustrated in Fig.~\ref{fig:framework}. The UE runs a lightweight local LLM for on-device inference, whereas the edge server maintains a pool of larger LLMs.

For each query, the system selects one model from a finite candidate set. Let
\begin{equation}
\mathcal{M} \triangleq \{m_0,m_1,\ldots,m_K\}
\end{equation}
denote the set of all candidate models, where $m_0$ is the lightweight model deployed on the UE and
\begin{equation}
\mathcal{M}_{\rm edge}\triangleq\mathcal{M}\setminus\{m_0\}
\end{equation}
is the set of eligible edge models. Although $\mathcal{M}_{\rm edge}$ is modeled as a single logical tier, different edge models may be instantiated on heterogeneous platforms. Routing selects one model from $\mathcal{M}$. 

The deployment architecture induces a two-stage routing process. The UE first decides whether to accept the query locally or defer it to the edge, and only deferred queries are subsequently routed among edge models. The UE can access only local information, including the query representation and an operator-specified accuracy-cost preference. It cannot access edge model utilities or any signals that require communication with the edge server. This restriction defines the key constraint of the routing problem: the UE must make the local acceptance decision without access to edge alternatives.

Let $\xi$ denote the runtime system state used in the deployment-cost model. $\xi$ captures stochastic wireless conditions, including uplink and downlink channel gains. These variations directly impact communication latency and energy consumption. All other quantities, such as bandwidth, transmit power, noise spectral density, and model-specific execution coefficients, are treated as fixed system parameters.

\subsection{Wireless Communication Model}
Let $L_{\rm in}(x)$ denote the input token count and let $L_{{\rm out},m}(x)$ denote the estimated output token count used for pre-execution routing. We assume that this estimate is accurate for instantiating the deployment-cost model. Let $b_{\rm in}$ and $b_{\rm out}$ denote the effective numbers of transmitted bits per input and output token. The uplink and downlink payload sizes are
\begin{equation}
S^{\rm ul}(x)=b_{\rm in}L_{\rm in}(x),
\end{equation}
\begin{equation}
S_m^{\rm dl}(x)=b_{\rm out}L_{{\rm out},m}(x).
\end{equation}

Since all edge models share the same wireless link, the transmission rate depends on the system state $\xi$. Under quasi-static fading, the uplink transmission rate is
\begin{equation}
R^{\rm ul}(\xi)
=
B^{\rm ul}
\log_2\!\left(
1+\frac{p_u^{\rm ul}\,g^{\rm ul}(\xi)}{B^{\rm ul}N_0}
\right),
\end{equation}
where $p_u^{\rm ul}$ is the UE radiated transmit power, $g^{\rm ul}(\xi)$ is the uplink channel gain incorporating small-scale fading and distance-dependent path loss, $B^{\rm ul}$ is the uplink bandwidth, and $N_0$ is the noise power spectral density. Similarly, the downlink transmission rate is
\begin{equation}
R^{\rm dl}(\xi)
=
B^{\rm dl}
\log_2\!\left(
1+\frac{p_{\rm bs}^{\rm dl}\,g^{\rm dl}(\xi)}{B^{\rm dl}N_0}
\right),
\end{equation}
where $p_{\rm bs}^{\rm dl}$ is the edge access-point radiated transmit power, $g^{\rm dl}(\xi)$ is the downlink channel gain, and $B^{\rm dl}$ is the downlink bandwidth. The channel gains can be parameterized as $g^q(\xi)=K_0 |z^q|^2(d_{\text{UE}}/d_0)^{-\alpha_{\rm pl}}$, $q\in\{\rm ul,dl\}$, where $d_{\text{UE}}$ is the UE--edge distance, $|z^q|^2$ is the fading power gain, $d_0=1$ m, and $K_0$ is the linear-scale reference gain.

The resulting communication delays are
\begin{equation}
\tau^{\rm ul}(x,\xi)=\frac{S^{\rm ul}(x)}{R^{\rm ul}(\xi)},
\qquad
\tau_m^{\rm dl}(x,\xi)=\frac{S_m^{\rm dl}(x)}{R^{\rm dl}(\xi)}.
\end{equation}

\subsection{Computation Model}

For any selected model $m\in\mathcal{M}$, the computation workload of serving query $x$ consists of a prompt-prefill stage and an autoregressive decoding stage. Let $\Phi_m(x)$ denote the total floating-point workload required by model $m$ on query $x$. Then
\begin{equation}
\Phi_m(x)
=
\Phi_m^{\rm pre}\!\bigl(L_{\rm in}(x)\bigr)
+
\Phi_m^{\rm dec}\!\bigl(L_{{\rm out},m}(x)\bigr),
\end{equation}
where $\Phi_m^{\rm pre}(\cdot)$ and $\Phi_m^{\rm dec}(\cdot)$ denote the prefill and decoding workload components, respectively. The exact form of these workload terms can be obtained from standard transformer complexity approximations or from direct profiling.

Let $\beta_m^{\rm pre}$ and $\beta_m^{\rm dec}$ denote the profiled per-operation execution times for the prefill and decoding stages of model $m$, respectively. These coefficients are fixed for each model and platform pair and are obtained by profiling each model on its designated hardware. The inference latency of model $m$ is then modeled as
\begin{equation}
\tau_m^{\rm inf}(x)
=
\beta_m^{\rm pre}\,\Phi_m^{\rm pre}\!\bigl(L_{\rm in}(x)\bigr)
+
\beta_m^{\rm dec}\,\Phi_m^{\rm dec}\!\bigl(L_{{\rm out},m}(x)\bigr).
\end{equation}

\subsection{End-to-End Latency Model}

The end-to-end latency of a selected model consists of communication delay and inference delay. For the local model $m_0$, only on-device inference is involved, and thus
\begin{equation}
t_0(x,\xi)=\tau_0^{\rm inf}(x).
\end{equation}

For an edge model $m\in\mathcal{M}_{\rm edge}$, the end-to-end latency includes uplink communication, a fixed UE--edge round-trip overhead, edge inference, and downlink communication. Let $\tau^{\rm rtt}$ capture protocol, access, and scheduling overhead that is not included in the payload transmission terms
$\tau^{\rm ul}$ and $\tau_m^{\rm dl}$. Therefore, we have
\begin{equation}
t_m(x,\xi)=\tau^{\rm ul}(x,\xi)+\tau^{\rm rtt}+\tau_m^{\rm inf}(x)+\tau_m^{\rm dl}(x,\xi), m\in\mathcal{M}_{\rm edge}.
\end{equation}

\subsection{Energy Model}

For the local model $m_0$, the energy consumption is entirely due to on-device inference:
\begin{equation}
e_0(x,\xi)=P_0^{\rm act}\, \tau_0^{\rm inf}(x),
\end{equation}
where $P_0^{\rm act}$ is the active power of UE during local inference.

For an edge model $m\in\mathcal{M}_{\rm edge}$, the total energy includes UE transmission, reception, and idle waiting, as well as edge-server computation:
\begin{equation}
\begin{aligned}
e_m(x,\xi)={} &
P_u^{\rm tx}\,\tau^{\rm ul}(x,\xi)
+ P_u^{\rm rx}\,\tau_m^{\rm dl}(x,\xi) \\
&+ P_u^{\rm idle}\bigl(\tau^{\rm rtt}+\tau_m^{\rm inf}(x)\bigr)
+ P_m^{\rm srv}\,\tau_m^{\rm inf}(x),
\end{aligned}
\end{equation}
where $P_u^{\rm tx}$ and $P_u^{\rm rx}$ are the UE power consumption during uplink transmission and downlink reception, respectively, $P_u^{\rm idle}$ is the UE idle power during edge inference, and $P_m^{\rm srv}$ is the active power of the edge server executing model $m$.

\subsection{Deployment Cost Model}
To combine latency and energy into a single deployment objective, we first define a dimensionless raw cost:
\begin{equation}
\bar c_m(x,\xi)
=
\omega_t \frac{t_m(x,\xi)}{T_0}
+
\omega_e \frac{e_m(x,\xi)}{E_0},
\qquad
\omega_t+\omega_e=1 .
\end{equation}
Here, $T_0$ and $E_0$ are fixed reference scales, computed as the mean latency and energy of the largest model on a reference set. They merely normalize latency and energy before weighted aggregation.

For comparable routing decisions, we further express each model cost relative to the same query reference model:
\begin{equation}
c_m(x,\xi)
=
\frac{\bar c_m(x,\xi)}
{\bar c_{m_{\rm ref}}(x,\xi)},
\end{equation}
where $m_{\rm ref}$ is the largest model in the pool. Therefore, $c_{m_{\rm ref}}(x,\xi)=1$ for every query--state pair by construction, and the normalized deployment cost of a routing policy is the sample average of the normalized costs of its selected models.

\section{Problem Formulation and Analysis}
\label{sec:problem}
\subsection{Problem Formulation}

Let $\mathcal{X}$ denote the query space and $\Xi$ the runtime system-state space. For each query--state pair $(x,\xi)\in\mathcal{X}\times\Xi$, collaborative routing selects one model from the candidate set $\mathcal{M}$ introduced in Sec.~\ref{sec:sysmodel}. Let $y_m(x)\in\{0,1\}$ denote the correctness indicator of model $m$ on query $x$, i.e.,
\begin{equation}
y_m(x)=
\begin{cases}
1, & \text{if model } m \text{ answers } x \text{ correctly},\\
0, & \text{otherwise}.
\end{cases}
\end{equation}

%Selecting model $m\in\mathcal{M}$ incurs a state-dependent deployment cost $c_m(x,\xi)$ as defined in Sec.~\ref{sec:sysmodel}.

A routing policy maps each query--state pair to one model:
\begin{equation}
\pi:\mathcal{X}\times\Xi\rightarrow\mathcal{M}.
\end{equation}

Let $\mathcal{D}$ denote the joint distribution of queries and runtime states. The induced routing accuracy and expected deployment cost are
\begin{equation}
\mathcal{A}(\pi)
=
\mathbb{E}_{(x,\xi)\sim\mathcal{D}}
\!\left[
y_{\pi(x,\xi)}(x)
\right],
\end{equation}
\begin{equation}
\mathcal{C}(\pi)
=
\mathbb{E}_{(x,\xi)\sim\mathcal{D}}
\!\left[
c_{\pi(x,\xi)}(x,\xi)
\right],
\end{equation}

The routing objective is
\begin{equation}
\begin{aligned}
\max_{\pi}\quad & \mathcal{A}(\pi)\\
\text{s.t.}\quad & \mathcal{C}(\pi)\le \Gamma,
\end{aligned}
\label{eq:population_problem}
\end{equation}
where $\Gamma>0$ is an operator-specified deployment-cost budget. However, the flat selector in~\eqref{eq:population_problem} does not encode the information structure of the mobile-edge deployment. The UE can access only local information; edge-model scores are unavailable unless the query is deferred. Hence, the routing policy cannot be an arbitrary selector over $\mathcal{M}$.

This leads to a deployment-constrained two-stage routing framework used in this paper, which restricts feasible policies to a factorized form. Let $o^{\rm dev}(x)\in\mathcal{O}_{\rm dev}$ denote the UE-observable information available at the UE for query $x$. A deployment-feasible routing policy is
\begin{equation}
\pi_\lambda(x,\xi)=
\begin{cases}
m_0, & a_\lambda\!\left(o^{\rm dev}(x)\right)=1,\\
\rho_\lambda(x,\xi), &
  a_\lambda\!\left(o^{\rm dev}(x)\right)=0,
\end{cases}
\label{eq:factorized_policy}
\end{equation}
where
$a_\lambda:\mathcal{O}_{\rm dev}\rightarrow\{0,1\}$
is the UE local-acceptance indicator, with
$a_\lambda(o^{\rm dev})=1$ denoting local execution and
$a_\lambda(o^{\rm dev})=0$ denoting deferral, and
$\rho_\lambda:\mathcal{X}\times\Xi\rightarrow
\mathcal{M}_{\rm edge}$ is the edge selector applied only to deferred queries. The factorization does not imply that the first-stage gate observes the full runtime state $\xi$; state dependence is retained in the realized costs and in the edge selector.

The deployment-constrained routing problem over the
factorized policy family is then
\begin{equation}
\begin{aligned}
\max_{\pi_\lambda\in\Pi_{\rm fac}}\quad &
  \mathcal{A}(\pi_\lambda)\\
\text{s.t.}\quad &
  \mathcal{C}(\pi_\lambda)\le \Gamma,
\end{aligned}
\label{eq:factorized_population_problem}
\end{equation}
where $\Pi_{\rm fac}$ denotes the family of factorized policies in~\eqref{eq:factorized_policy}.

To parameterize different accuracy--cost operating points, we use the scalarized utility
\begin{equation}
u_m(x,\xi;\lambda)=y_m(x)-\lambda\, c_m(x,\xi),
\qquad m\in\mathcal{M},
\label{eq:utility}
\end{equation}
where $\lambda\ge0$ is a cost weight selected by the operator.
Larger $\lambda$ values place more weight on deployment cost and therefore favor lower-cost models.
For any policy $\pi$, this utility satisfies
\begin{equation}
    \mathbb{E}_{\mathcal{D}}\!\left[
    u_{\pi(x,\xi)}(x,\xi;\lambda)
    \right]
    =
    \mathcal{A}(\pi)-\lambda\mathcal{C}(\pi).
\end{equation}

Thus, the scalarized objective provides a practical way to parameterize accuracy--cost trade-offs via a single weight $\lambda$. By varying $\lambda$, we obtain a family of routing policies that trace different operating points on the accuracy--cost frontier. In practice, we optimize and calibrate policies over a range of $\lambda$ values and evaluate their realized performance, rather than solving separate constrained problems for fixed budgets $\Gamma$.

Given a finite sample
$\mathcal{S}=\{(x_i,\xi_i)\}_{i=1}^{N}$ with model-wise
correctness annotations
$\{y_m(x_i)\}_{m\in\mathcal{M}}$, the empirical
budgeted counterpart of~\eqref{eq:factorized_population_problem} is
\begin{equation}
\begin{aligned}
\max_{\pi_\lambda\in\Pi_{\rm fac}}\quad &
\frac{1}{N}\sum_{i=1}^{N}
y_{\pi_\lambda(x_i,\xi_i)}(x_i)\\
\text{s.t.}\quad &
\frac{1}{N}\sum_{i=1}^{N}
c_{\pi_\lambda(x_i,\xi_i)}(x_i,\xi_i)
\le \Gamma.
\end{aligned}
\label{eq:factorized_empirical_problem}
\end{equation}

\subsection{Design Challenges of the Factorized Formulation}

Although~\eqref{eq:factorized_population_problem} captures the deployment structure of the mobile-edge system, it reveals three key challenges in realizing a deployable routing policy. To analyze the first-stage decision induced by the factorized policy, we define the full-information local-versus-edge utility margin
\begin{equation}
\Delta(x,\xi;\lambda)
=
u_{m_0}(x,\xi;\lambda)
-
\max_{m\in\mathcal{M}_{\rm edge}} u_m(x,\xi;\lambda).
\label{eq:reference_margin}
\end{equation}
The sign of $\Delta(x,\xi;\lambda)$ characterizes the full-information
local-versus-edge preference at operating point $\lambda$:
$\Delta\ge0$ indicates that local execution has no smaller scalarized
utility than any edge alternative, whereas $\Delta<0$ indicates that
an edge model is preferred. This margin serves as a reference object
for analyzing the first-stage local acceptance decision.

\paragraph{Partial-information local acceptance}
The UE must decide whether to accept locally or defer using only locally observable signals, whereas the full-information preference in \eqref{eq:reference_margin} depends on edge model utilities and the runtime state. Moreover, correctness is driven primarily by the semantic difficulty of $x$, while deployment cost also depends on $\xi$. This mismatch prevents direct deployment of full-information decision rules and rules out standard model-selection classifiers. The first-stage gate must therefore learn a $\lambda$-conditioned surrogate of the local-versus-edge preference using only query representations.

\paragraph{Asymmetric local-defer errors}
A false local acceptance misses escalation to a preferred edge model and degrades user-perceived quality, whereas an unnecessary deferral mainly increases latency and energy cost. These error modes are asymmetric and cannot be adequately controlled by optimizing average accuracy and cost. The local decision rule should therefore be conservative about local acceptance and expose a mechanism to control the marginal false-acceptance risk at each operating point. This necessitates a calibration step that maps the learned score to a deployable threshold with explicit risk control.

\paragraph{State-dependent deferred selection}
Deferral does not eliminate cost sensitivity. Wireless delay, edge execution latency, and energy consumption vary with $\xi$, making the relative utility of edge models inherently state-dependent. Ignoring state variation would lead to suboptimal routing decisions. The deferred branch must therefore incorporate system state to maintain cost-aware model selection.

Together, these requirements motivate CR$^2$: a UE-observable margin
surrogate, state-aware deferred selection, and CRC-calibrated local
acceptance.

\section{The proposed CR$^2$ scheme}
\label{sec:method}

Within the deployment-constrained factorized routing framework in Sec.~\ref{sec:problem}, the goal is to realize a deployable routing policy that handles the three challenges, while preserving the cost-aware utility structure induced by the deployment cost model in Sec.~\ref{sec:sysmodel}.

\begin{figure*}[t]
  \centering
  \includegraphics[width=\textwidth]{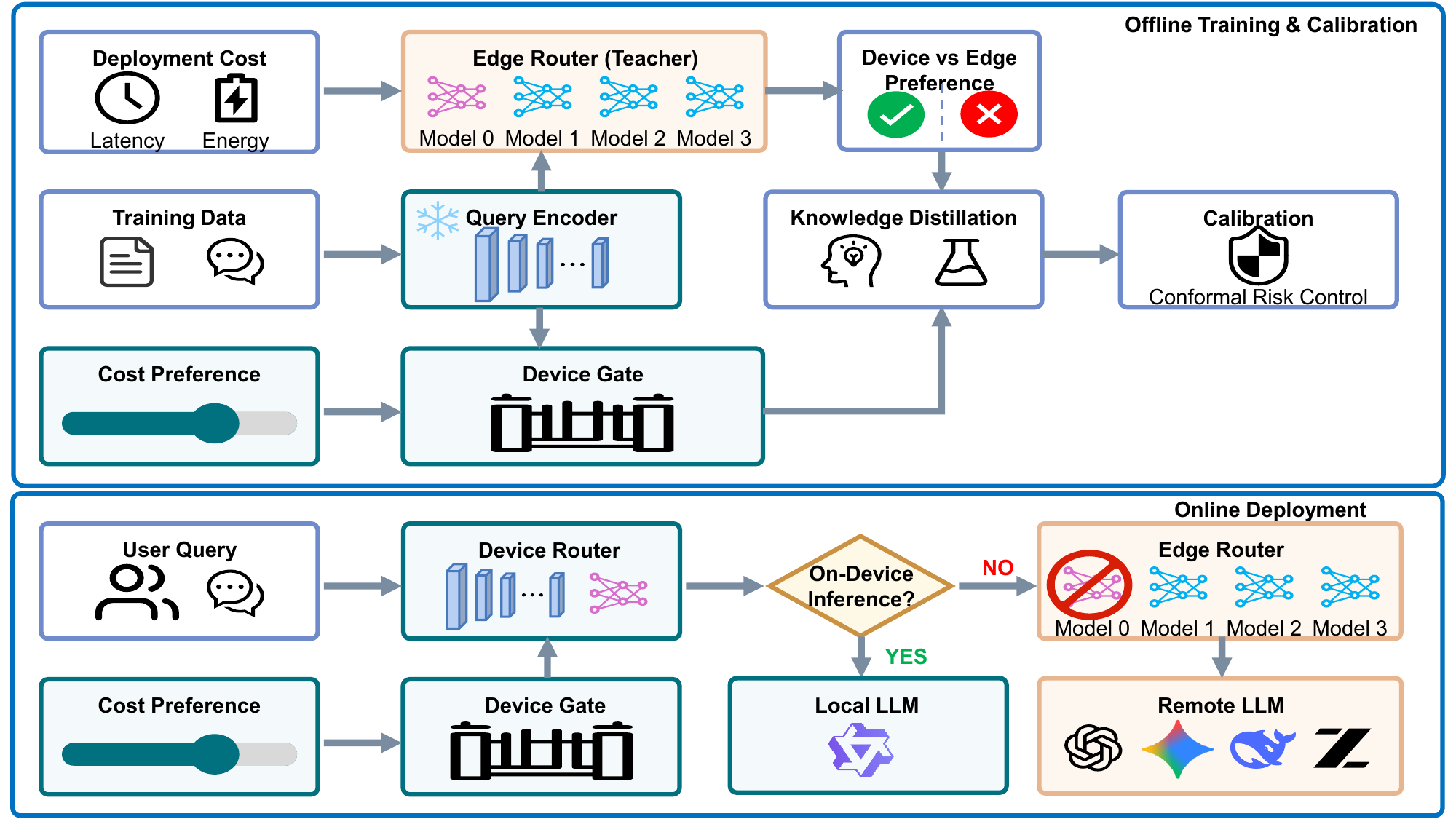}
\caption{Overview of CR$^2$ including offline training, CRC-based calibration, and online two-stage routing.}
  \label{fig:cr2_methodology}
\end{figure*}

The proposed CR$^2$ addresses these challenges through three coupled components: a lightweight margin gate for partial-information local acceptance, a teacher selector that provides full-information utility margins during training and whose utility formulation is reused by the edge router at inference time after deferral, and a per-$\lambda$ CRC calibration procedure that converts the gate score into a calibrated acceptance rule. Each component corresponds to one challenge: the margin gate enables partial-information decision making at the UE, CRC provides risk-controlled local acceptance, and the teacher/edge selector preserves state-aware utility optimization for deferred queries. These components have distinct statistical roles: the teacher and margin gate are learned from finite training data and then treated as fixed scoring functions for calibration, whereas CRC calibrates the local acceptance threshold of that fixed score against the routing-specific risk defined below. Fig.~\ref{fig:cr2_methodology} summarizes the resulting offline training, calibration, and online routing workflow.

\subsection{Two-Component Decision Rule}
\label{sec:method_rule}

For each query $x$, let
\begin{equation}
e(x)=f_{\rm emb}(x)\in\mathbb{R}^{d}
\end{equation}
denote the output of a frozen query encoder $f_{\rm emb}$. The teacher selector $f_\theta$ produces per-model correctness estimates
\begin{equation}
\{p_m(x)\}_{m\in\mathcal{M}}\subset[0,1].
\end{equation}
In CR$^2$, the UE-observable argument in \eqref{eq:factorized_policy} is instantiated by the local query embedding together with the operator-selected cost weight; equivalently, the learned first-stage gate is evaluated on
$o^{\rm dev}_\lambda(x)=(e(x),\lambda)$.

For a runtime state $\xi$ and a cost weight
$\lambda$, we define the teacher-estimated utility
\begin{equation}
\hat u_m(x,\xi;\lambda)
=
p_m(x)-\lambda c_m(x,\xi),
\qquad m\in\mathcal{M}.
\label{eq:teacher_utility}
\end{equation}

The margin gate does not observe $\xi$ or edge-model utility estimates before deferral. These quantities are used offline for teacher construction and online only by the edge router after deferral.

For deferred queries, the edge router performs utility maximization over the candidate set $\mathcal{M}_{\rm edge}$. The deferred selection rule is
\begin{equation}
m_{\rm def}^\star(x,\xi;\lambda)
=
\arg\max_{m\in\mathcal{M}_{\rm edge}}
\hat u_m(x,\xi;\lambda).
\label{eq:deferred_argmax}
\end{equation}

Two useful deployment variants are: 1) the inclusive variant
$\mathcal{M}_{\rm def}=\mathcal{M}$, in which deferred queries may still be mapped back to the local model if it remains utility-optimal; and 2) the edge-only variant $\mathcal{M}_{\rm def}=\mathcal{M}_{\rm edge}$,
in which deferral commits the query to edge execution.

The device predicts a local-versus-edge utility margin from
$(e(x),\lambda)$ only. Let
$\hat\Delta_\phi(e(x),\lambda)\in\mathbb{R}$ denote the raw output of the
margin gate, and define the corresponding score
\begin{equation}
s_\phi(e(x),\lambda)
=
\sigma\!\left(
\frac{\hat\Delta_\phi(e(x),\lambda)}{T}
\right)\in[0,1],
\label{eq:gate_score}
\end{equation}
where $T>0$ is a learnable temperature. Let
$\tau^\star(\lambda;\alpha)$ denote the calibrated local acceptance
threshold for target calibration level $\alpha$. The deployed routing
rule is
\begin{equation}
\hat m(x,\xi;\lambda)=
\left\{
\begin{array}{ll}
m_0, & s_\phi(e(x),\lambda)\ge \tau^\star(\lambda;\alpha),\\
m_{\rm def}^\star(x,\xi;\lambda), & \text{otherwise}.
\end{array}
\right.
\label{eq:cr2_rule}
\end{equation}

\subsection{Teacher Selector}
\label{sec:teacher}

The teacher selector $f_\theta$ uses one binary MLP head per candidate model on top of the same frozen embedding used by the gate. Given $e(x)$, the selector produces one logit $\ell_m(x)\in\mathbb{R}$ for each model $m\in\mathcal{M}$, and the corresponding correctness estimate is
\begin{equation}
p_m(x)=\sigma(\ell_m(x)).
\end{equation}
These estimates are used in the teacher-estimated utility model in~\eqref{eq:teacher_utility}. The selector is trained once on the training split from correctness labels, frozen before gate fitting, and then reused unchanged on the validation and test splits.

The teacher selector is trained with both per-model binary cross-entropy and an intra-query pairwise ranking term. The ranking term encourages models that answer a query correctly to receive larger logits than models that fail on the same query. For a mini-batch $\mathcal{B}$, the teacher classification loss is
\begin{equation}
\mathcal{L}_{\rm BCE}^{\rm teacher}
=
\frac{1}{|\mathcal{B}|\,|\mathcal{M}|}
\sum_{i\in\mathcal{B}}\sum_{m\in\mathcal{M}}
\ell_{\rm BCE}\!\left(
\ell_m(x_i),y_m(x_i)
\right),
\label{eq:teacher_bce_loss}
\end{equation}
where the binary cross-entropy loss with logits is defined as
\begin{equation}
\ell_{\rm BCE}(\ell, y)
=
-\, y \log \sigma(\ell)
- (1-y)\log \bigl(1-\sigma(\ell)\bigr).
\end{equation}
For each query, define
\begin{equation}
P_i^+=\{m\in\mathcal{M}: y_m(x_i)=1\},
\end{equation}
\begin{equation}
P_i^-=\{m\in\mathcal{M}: y_m(x_i)=0\}.
\end{equation}

The ranking term is applied only to queries for which both correct and
incorrect models exist:
\begin{equation}
\mathcal{B}'=\{i\in\mathcal{B}: P_i^+\neq\emptyset,\ P_i^-\neq\emptyset\}.
\end{equation}

Queries for which all models are correct or all models are incorrect are skipped by this ranking term. For each query $i\in\mathcal{B}'$, define the set of correct--incorrect model pairs as
\begin{equation}
\mathcal{P}_i = P_i^+\times P_i^- .
\end{equation}

For $\mathcal{B}'\neq\emptyset$, the per-query ranking loss is
\begin{equation}
\ell_i^{\rm rank}
=
\frac{1}{|\mathcal{P}_i|}
\sum_{(m^+,m^-)\in\mathcal{P}_i}
\operatorname{softplus}
\!\left(
\ell_{m^-}(x_i)-\ell_{m^+}(x_i)
\right),
\label{eq:teacher_rank_per_query}
\end{equation}
where $\operatorname{softplus}(x)=\log(1+e^x)$ ensures positivity. The pairwise ranking loss is
\begin{equation}
\mathcal{L}_{\rm rank}
=
\frac{1}{|\mathcal{B}'|}
\sum_{i\in\mathcal{B}'}
\ell_i^{\rm rank}.
\label{eq:teacher_rank_loss}
\end{equation}

If $\mathcal{B}'=\emptyset$, we set $\mathcal{L}_{\rm rank}=0$ for that
mini-batch. The teacher training objective is
\begin{equation}
\mathcal{L}_{\rm teacher}
=
w_{\rm cls}\mathcal{L}_{\rm BCE}^{\rm teacher}
+
w_{\rm rank}\mathcal{L}_{\rm rank}.
\label{eq:teacher_loss}
\end{equation}
Importantly, this ranking term is applied to the teacher logits $\ell_m(x)$ and does not involve deployment cost or the operating-point weight $\lambda$; cost-awareness enters only later through the utility
$\hat u_m(x,\xi;\lambda)=p_m(x)-\lambda c_m(x,\xi)$.

The teacher selector serves two roles. Over the full model set $\mathcal{M}$, it defines the full-information reference used for teacher construction and comparison. Applied only after deferral over $\mathcal{M}_{\rm def}$, it determines the edge second-stage decision in the deployable realization. This separation keeps the margin gate anchored to a fixed teacher while ensuring that no edge-model utility estimate is exposed to the UE before deferral.

\subsection{Utility Margin and Teacher Signal}
\label{sec:margin}
The gate is trained to predict the local-versus-edge utility gap induced by the teacher-estimated utility in~\eqref{eq:teacher_utility}. This gate-level utility margin is distinct from the pairwise ranking loss used to train the teacher selector in~\eqref{eq:teacher_rank_loss}: the former incorporates deployment cost and the operating point $\lambda$, whereas the latter only enforces intra-query ranking of correctness logits. The full-information teacher-induced decision rule as
\begin{equation}
m_{\rm full}^\star(x,\xi;\lambda)
=
\arg\max_{m\in\mathcal{M}}
\hat u_m(x,\xi;\lambda).
\end{equation}
When multiple models attain the same teacher-estimated utility, the full-information reference selects the one with the smaller deployment cost.

The corresponding teacher utility margin is
\begin{equation}
\Delta_\theta(x,\xi;\lambda)
=
\hat u_{m_0}(x,\xi;\lambda)
-
\max_{m\in\mathcal{M}_{\rm edge}}
\hat u_m(x,\xi;\lambda).
\label{eq:teacher_margin}
\end{equation}
Ties involving the local model $m_0$ are broken in favor of $m_0$.

The sign of $\Delta_\theta$ indicates whether local execution is optimal under the teacher utility, while its magnitude reflects the strength of preference between local and edge execution. This margin serves as an offline surrogate of the oracle margin in~\eqref{eq:reference_margin}. We also define the corresponding binary label as
\begin{equation}
z(x,\xi;\lambda)
=
\mathbf{1}\!\bigl[\Delta_\theta(x,\xi;\lambda)\ge 0\bigr].
\label{eq:sign_label}
\end{equation}

The label $z$ captures the decision boundary between local acceptance and deferral, whereas the magnitude of $\Delta_\theta$ provides a continuous measure of decision confidence. This continuous structure is used to stabilize learning across different operating points indexed by $\lambda$ and to support subsequent threshold calibration.

We further assume that the normalized deployment cost is such that the local model is no more expensive than any edge model. Under this assumption, $\Delta_\theta(x,\xi;\lambda)$ is non-decreasing in $\lambda$ for fixed $(x,\xi)$, which provides a structural monotonicity prior for gate learning and calibration.

Importantly, the margin in~\eqref{eq:teacher_margin} compares local execution against the best edge alternative and is independent of whether the deferred execution uses the inclusive set $\mathcal{M}_{\rm def}=\mathcal{M}$ or the edge-only set $\mathcal{M}_{\rm def}=\mathcal{M}_{\rm edge}$. It therefore captures the operational event of interest at the UE: whether local acceptance would incur a loss relative to the best available edge model under the teacher utility.

\subsection{On-Device Margin Gate}
\label{sec:gate}

The gate predicts a real-valued approximation of the teacher margin:
\begin{equation}
\hat\Delta_\phi(e(x),\lambda)\in\mathbb{R}.
\end{equation}

To enable a single model to generalize across different accuracy--cost operating points, we explicitly condition the gate on the cost weight $\lambda$. Since $\lambda$ is strictly positive, we operate in logarithmic space and define a periodic feature mapping over
$\log\lambda$:
\begin{equation}
\psi_\lambda=
\Bigl[
\log\lambda,\,
\{\sin(2\pi f_k\log\lambda),\cos(2\pi f_k\log\lambda)\}_{k=1}^{4}
\Bigr],
\label{eq:psi_lambda}
\end{equation}
with frequencies $\{f_k\}=\{0.5,1,2,4\}$. The embedding branch computes
\begin{equation}
h=W_1\,\operatorname{LN}(e),
\end{equation}
while the cost-weight branch outputs the feature-wise affine parameters
\begin{equation}
(\gamma,\beta)=W_{\rm film}\,\psi_\lambda.
\end{equation}
Then, we have
\begin{equation}
\tilde h=\gamma\odot h+\beta,
\end{equation}
\begin{equation}
\hat\Delta_\phi
=
W_2\,\operatorname{Drop}\!\bigl(\operatorname{GELU}(\tilde h)\bigr)+b_2,
\label{eq:film_gate}
\end{equation}
where $\operatorname{GELU}(\cdot)$ denotes the Gaussian error linear unit activation and $\operatorname{Drop}(\cdot)$ denotes dropout regularization applied during training.
This structure allows the operating point $\lambda$ to modulate the
query representation directly in the hidden feature space.

\subsubsection{Temperature-scaled score}
The raw gate output is converted into the score used for calibration through a learnable positive temperature
\begin{equation}
T=\operatorname{softplus}(\eta_T)+10^{-6}.
\label{eq:temperature}
\end{equation}
The resulting score is exactly the quantity in~\eqref{eq:gate_score}. Learning $T$ jointly with the gate allows the distilled margin scale and the thresholding scale to be decoupled.

\subsubsection{Deployment footprint.} The deployed margin gate consists of one frozen embedding forward pass and one small MLP evaluation before the threshold lookup. It does not estimate an opaque confidence for local execution, but instead predicts how competitive local execution is relative to the best edge alternative.

\subsection{Margin-Distillation Objective}
\label{sec:gate_loss}

For each mini-batch, we independently sample
\begin{equation}
\{\lambda_j\}_{j=1}^{J}
\quad \text{with} \quad
\log \lambda_j \stackrel{\rm iid}{\sim} \mathcal{U}(\log 0.1,\log 20),
\end{equation}
where $J$ is the number of sampled operating points per mini-batch.
With batch size $B_{\rm mb}$, we obtain teacher margins
\(
\Delta_\theta\in\mathbb{R}^{J\times B_{\rm mb}}
\),
sign labels
\(
z\in\{0,1\}^{J\times B_{\rm mb}}
\),
and predicted margins
\(
\hat\Delta\in\mathbb{R}^{J\times B_{\rm mb}}
\).
The total training objective is
\begin{equation}
\mathcal{L}
=
w_{\rm sign}\,\mathcal{L}_{\rm sign}
+
w_{\rm margin}\,\mathcal{L}_{\rm margin}
+
w_{\rm mono}\,\mathcal{L}_{\rm mono}.
\label{eq:loss_total}
\end{equation}

\paragraph{Sign loss.}
The first term aligns the sign of the gate output with the local/defer
boundary induced by the teacher:
\begin{equation}
\mathcal{L}_{\rm sign}
=
\frac{1}{J B_{\rm mb}}\sum_{j,i}
\ell_{\rm BCE}\!\Bigl(
\hat\Delta_{j,i}/T,\,
z_{j,i}
\Bigr).
\label{eq:loss_bce}
\end{equation}

\paragraph{Margin loss.}
The second term regresses the raw gate output toward the continuous teacher margin. Specifically, we minimize a Huber loss on the residual between the predicted margin and the teacher margin:
\begin{equation}
\mathcal{L}_{\rm margin}
=
\frac{1}{J B_{\rm mb}}\sum_{j,i}
\ell_{\rm Huber}^{\beta}
\!\Bigl(
\hat\Delta_{j,i}
-
\operatorname{sg}(\Delta_{\theta,j,i})
\Bigr),
\label{eq:loss_reg}
\end{equation}
where $\operatorname{sg}(\cdot)$ denotes the stop-gradient operator,
i.e., $\frac{\partial}{\partial x}\operatorname{sg}(x)=0$, and
$\ell_{\rm Huber}^{\beta}(\cdot)$ is the Huber loss with transition
parameter $\beta>0$, defined as
\begin{equation}
\ell_{\rm Huber}^{\beta}(r)
=
\begin{cases}
\frac{1}{2} r^2, & |r|\le \beta,\\[1mm]
\beta\bigl(|r|-\frac{1}{2}\beta\bigr), & |r|>\beta.
\end{cases}
\end{equation}

This loss preserves local geometry of the teacher margin in its quadratic region while reducing sensitivity to large residuals via its linear regime. Although $\Delta_{\theta,j,i}$ depends on sampled runtime state $\xi_i$, the gate input only uses $(e(x_i),\lambda_j)$. Thus, the same query--weight pair may yield different teacher margins under different runtime states. The regression term learns a state-marginal surrogate of the full-information teacher margin under the deployment-state distribution. CRC-based calibration then converts the score into a deployable acceptance rule on held-out samples.

\paragraph{Monotone loss.}
Since the teacher margin is non-decreasing in $\lambda$ under the
profiled-regime assumption above, the learned margin should satisfy the
same structural property. Let
$\lambda_{(1)}<\cdots<\lambda_{(J)}$ denote the sorted sampled cost
weights, and let $\tilde\Delta$ denote the corresponding re-ordered
predicted margins. The monotonicity penalty is
\begin{equation}
\mathcal{L}_{\rm mono}
=
\frac{1}{(J-1)B_{\rm mb}}
\sum_{j=1}^{J-1}\sum_i
\bigl(
\tilde\Delta_{j,i}-\tilde\Delta_{j+1,i}
\bigr)_+,
\label{eq:loss_mono}
\end{equation}
where $(x)_+ \triangleq \max\{0,x\}$.

\paragraph{Training protocol.}The gate parameters and temperature are optimized jointly. The selected checkpoint is frozen before computing the threshold table. Thus, the learned score is evaluated as a supervised routing model on held-out queries, while CRC is used only to calibrate the marginal false-acceptance risk of the fixed score, rather than to provide a generalization bound for the teacher or gate. Loss weights and implementation hyperparameters are reported in Sec.~\ref{sec:experiments}.

\subsection{Per-$\lambda$ CRC Calibration of the Acceptance Threshold}
\label{sec:crc}

The score $s_\phi(e,\lambda)$ is not itself an interpretable acceptance criterion, because its scale may vary with the operating point $\lambda$. We therefore calibrate a $\lambda$-indexed threshold family $\tau^\star(\lambda;\alpha)$ using CRC applied separately at each fixed cost weight.

\subsubsection{Marginal false-acceptance risk}

Fix a calibration set
$\{(x_i,\xi_i)\}_{i=1}^{N_v}$, drawn from the same query and runtime-state
sampling process as the future requests to which the threshold table will
be applied. This is the routing-specific exchangeability condition
corresponding to the CRC statement in Sec.~\ref{sec:preliminaries}; if
the query mixture or deployment-state distribution changes, the
thresholds should be recalibrated. For each $\lambda$, let
\begin{equation}
s_i^{(\lambda)}=s_\phi(e(x_i),\lambda)
\end{equation}
be the gate score, and define the teacher disagreement label
\begin{equation}
r_i^{(\lambda)}
=
\mathbf{1}\!\bigl[
m_{\rm full}^\star(x_i,\xi_i;\lambda)\neq m_0
\bigr].
\label{eq:disagree_label}
\end{equation}
Here $r_i^{(\lambda)}=1$ indicates that the full-information reference,
with cost-based tie-breaking, prefers an edge model over local execution. For any candidate threshold
$\tau$, define the routing-specific per-sample loss
\begin{equation}
\ell_i(\tau,\lambda)
=
r_i^{(\lambda)}
\mathbf{1}\!\bigl[
s_i^{(\lambda)}\ge \tau
\bigr].
\label{eq:marginal_loss}
\end{equation}
It equals one only when the gate accepts the query locally while the
teacher prefers edge execution. The empirical marginal false-acceptance
risk is
\begin{equation}
\hat R_{N_v}(\tau,\lambda)
=
\frac{1}{N_v}
\sum_{i=1}^{N_v}
r_i^{(\lambda)}
\mathbf{1}\!\bigl[
s_i^{(\lambda)}\ge \tau
\bigr].
\label{eq:rhat_tau}
\end{equation}

This is the calibration target used in this work. That is,
$\hat R_{N_v}(\tau,\lambda)$ measures the fraction of all calibration
queries that would be accepted locally even though the teacher-preferred
selection is an edge model. Unlike a quantity normalized only over accepted
queries, this quantity is normalized by the full calibration set size
and therefore matches the bounded monotone-loss form reviewed in
Sec.~\ref{sec:preliminaries}.

\begin{table}[!t]
	\centering
	\caption{System parameters.}
	\label{tab:sim_params}
	\resizebox{0.48\textwidth}{!}{%
		\begin{tabular}{lrl}
			\toprule
			\textbf{Parameter} & \textbf{Value} & \textbf{Description} \\
			\midrule
			\multicolumn{3}{l}{\emph{Communication}} \\
			$b_{\mathrm{in}},\, b_{\mathrm{out}}$ & 32 bits/token & Fixed effective bits per token \\
			$B^{\mathrm{ul}},\, B^{\mathrm{dl}}$ & 10, 40 MHz & Uplink/downlink bandwidths \\
			$\tau^{\mathrm{rtt}}$ & 18 ms & UE--edge round-trip overhead \\
			$f_c$ & 3.5 GHz & Carrier frequency \\
			$K_0$ & $-43.3$ dB at $d_0=1$ m & Free-space path-loss reference \\
			$d$ & $\mathcal{U}(30,150)$ m & UE--edge propagation distance \\
			$\alpha_{\rm pl}$ & 4.0 & Path-loss exponent (urban NLOS) \\
			$|h^{\mathrm{ul}}|^2,\, |h^{\mathrm{dl}}|^2$ & $\mathrm{Exp}(1)$ & Quasi-static fading power gains \\
			$p_u^{\mathrm{ul}},\, p_{\mathrm{bs}}^{\mathrm{dl}}$ & 0.5, 2.0 W & Radiated UE / edge access-point transmit powers \\
			$N_0$ & $-174$ dBm/Hz & Noise power spectral density \\
			\midrule
			\multicolumn{3}{l}{\emph{UE Power}} \\
			$P_u^{\mathrm{tx}},\, P_u^{\mathrm{rx}}$ & 1.20, 0.90 W & UE transmit/receive power consumption \\
			$P_u^{\mathrm{idle}}$ & 0.05 W & UE idle power \\
			$P_0^{\mathrm{act}}$ & 15.0 W & UE local-inference power \\
			\midrule
			\multicolumn{3}{l}{\emph{Edge Pool Power}} \\
			$P_m^{\mathrm{srv}}$ & 150, 300, 200 W & Edge active powers for Qwen3-4B/8B/14B \\
			\midrule
			\multicolumn{3}{l}{\emph{Deployment Cost}} \\
			$\omega_t,\,\omega_e$ & 0.5, 0.5 & Latency/energy weights \\
			$T_0$ & 1118.617 ms & Latency normalization constant \\
			$E_0$ & 223752.133 mJ & Energy normalization constant \\
			\bottomrule
	\end{tabular}}
\end{table}

\subsection{Online Inference Procedure}
\label{sec:training_online}
\begin{algorithm}[t]
	\caption{Online inference of CR$^2$}
	\label{alg:online_inference}
	\begin{algorithmic}[1]
		\REQUIRE Query $x$, operating parameters $(\alpha,\lambda)$, frozen
		embedding model $f_{\rm emb}$, gate $\hat\Delta_\phi$, score function
		$s_\phi$, calibrated threshold table $\tau^\star(\cdot\,;\cdot)$, and
		edge-side runtime state $\xi$ if the query is deferred.
		\ENSURE Selected execution model.
		\STATE Compute the UE-observable embedding $e(x)=f_{\rm emb}(x)$.
		\STATE Evaluate the predicted margin
		$\hat\Delta_\phi(e(x),\lambda)$ and acceptance score
		$s_\phi(e(x),\lambda)$.
		\STATE Retrieve the calibrated threshold
		$\tau^\star(\lambda;\alpha)$ from the shipped table.
		\IF{$s_\phi(e(x),\lambda)\ge \tau^\star(\lambda;\alpha)$}
		\STATE Execute the query locally using $m_0$.
		\ELSE
		\STATE Defer the query to the edge server.
		\STATE Select the edge model by the utility rule in~\eqref{eq:deferred_argmax}.
		\ENDIF
	\end{algorithmic}
\end{algorithm}

\subsubsection{CRC calibration}

For each fixed $(\lambda,\alpha)$, we evaluate candidate thresholds and
choose the least conservative threshold whose CRC correction is at most
$\alpha$. Since~\eqref{eq:marginal_loss} is binary and non-increasing in
$\tau$, the CRC correction at fixed $\lambda$ is
\begin{equation}
\operatorname{CRC}(\tau,\lambda)
=
\frac{N_v}{N_v+1}\hat R_{N_v}(\tau,\lambda)
+
\frac{1}{N_v+1}.
\label{eq:crc_tau}
\end{equation}
Equivalently, let $d_\tau(\lambda)$ be the number of calibration
queries satisfying both $s_i^{(\lambda)}\ge \tau$ and
$r_i^{(\lambda)}=1$, then
$\operatorname{CRC}(\tau,\lambda)=(d_\tau(\lambda)+1)/(N_v+1)$.
The calibrated threshold is
\begin{equation}
\tau^\star(\lambda;\alpha)
=
\inf\Bigl\{
\tau:\,
\operatorname{CRC}(\tau,\lambda)
\le \alpha
\Bigr\}.
\label{eq:tau_star}
\end{equation}
Under the exchangeability condition stated above and with the learned
score fixed before calibration, this rule controls the marginal
false-acceptance probability for a future request at the selected cost
weight:
\begin{equation}
\mathbb{E}\!\left[
r_{N_v+1}^{(\lambda)}
\mathbf{1}\!\bigl[
s_{N_v+1}^{(\lambda)}\ge \tau^\star(\lambda;\alpha)
\bigr]
\right]
\le \alpha .
\label{eq:marginal_crc_guarantee}
\end{equation}
The guarantee is pointwise for a pre-specified $\lambda$ and a fixed
learned score. It is not a simultaneous guarantee over the entire
$\lambda$ sweep.
This rule applies CRC to select a local acceptance threshold at a fixed
cost weight $\lambda$ rather than to calibrate the cost weight itself.

Given deployment parameters $(\alpha,\lambda)$, CR$^2$ applies the online procedure in Algorithm \ref{alg:online_inference}. Before deferral, the UE uses only $(e(x),\lambda)$
and a calibrated threshold table, and no
edge-model utility estimate is exposed. This is how the UE and edge server are decoupled in CR$^2$.

\begin{figure}[!t]
    \centering
    \subfloat[]{%
      \includegraphics[width=0.85\columnwidth]{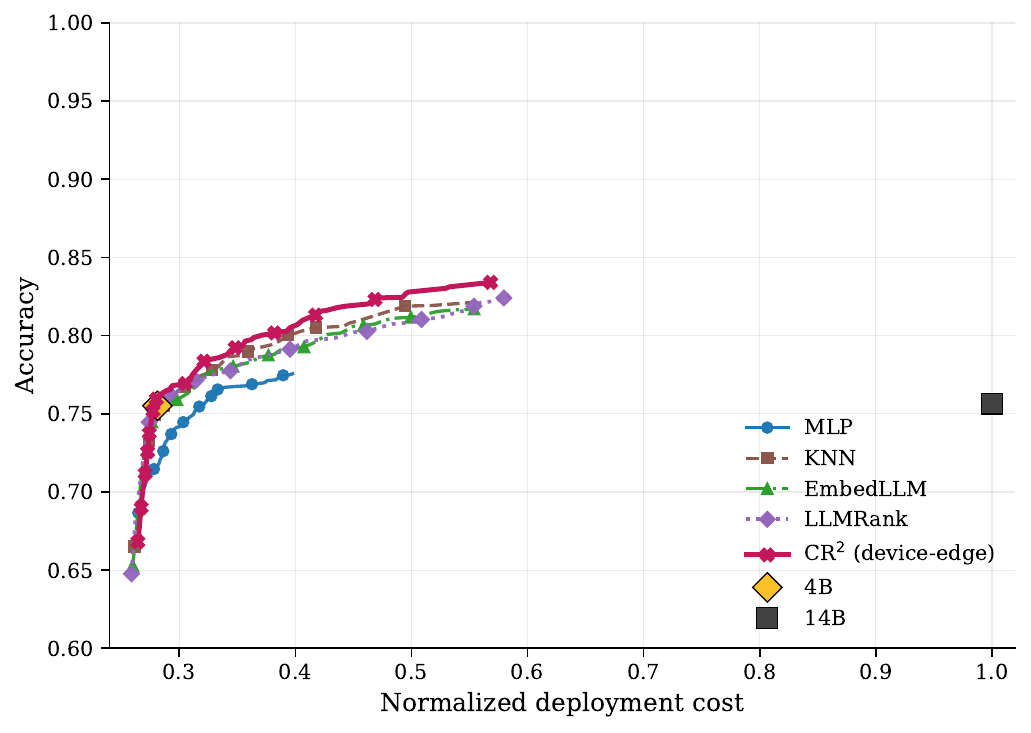}%
      \label{fig:pareto_full}%
    }\\[-0.5mm]
    \subfloat[]{%
      \includegraphics[width=0.85\columnwidth]{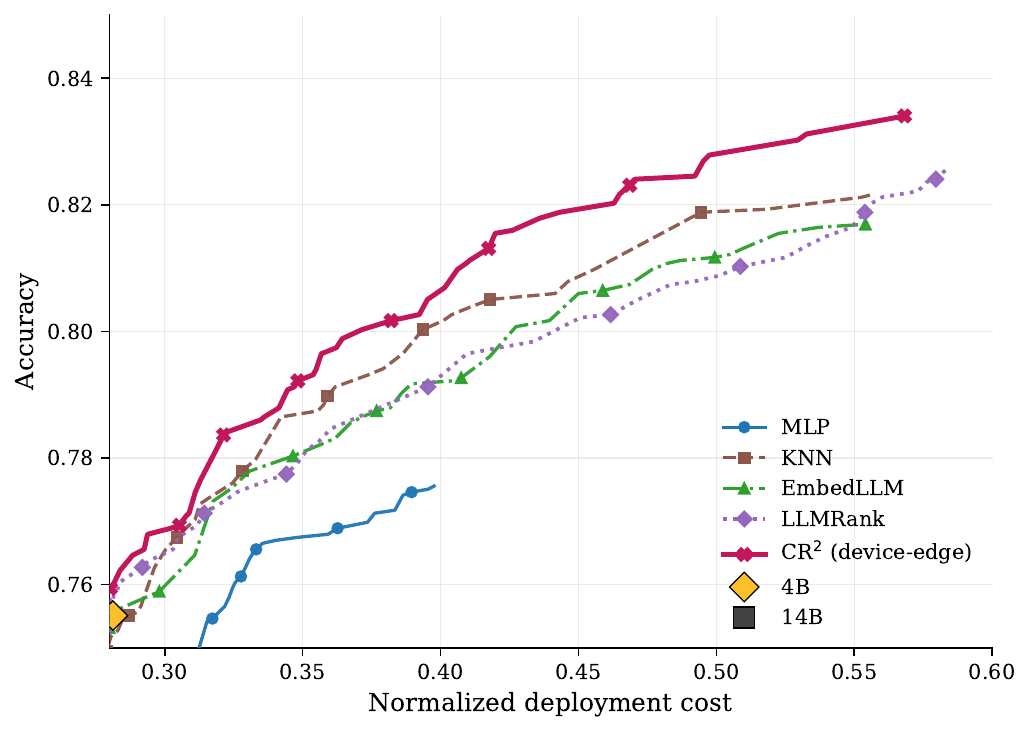}%
      \label{fig:pareto}%
    }
    \caption{Accuracy--cost Pareto curves: (a) full range and (b) zoomed operating region.}
    \label{fig:pareto_main}
\end{figure}

\section{Experiments}
\label{sec:experiments}

\subsection{Experimental Setup}
\label{sec:exp_setup}

Table~\ref{tab:sim_params} lists the communication, computation, and energy parameters used in the deployment-cost model. We instantiate the model pool with Qwen3-$1.7$B, Qwen3-$4$B, Qwen3-$8$B, and Qwen3-$14$B. Qwen3-$1.7$B is deployed on the UE using a Jetson AGX Orin profile, while the larger models run on RTX 4070 Ti, RTX 4090, and A40 GPUs at the edge. We profile inference latency and active power.

We construct a routing dataset from MMLU~\cite{hendrycks2021mmlu}, BBH~\cite{suzgun2022bbh}, GPQA~\cite{rein2023gpqa}, and MBPP~\cite{austin2021mbpp}, covering world knowledge, multi-step reasoning, graduate-level science, and code generation. Model-wise correctness labels are obtained using \texttt{lm-evaluation-harness}~\cite{gao2024harness}, following the dataset construction setting of EmbedLLM~\cite{zhuang2025embedllm}. The validation split is used for operating-point calibration, and the test split is held out for routing and calibration-risk evaluation under the same runtime-state sampling protocol. Test queries where no model answers correctly are excluded.

We compare CR$^2$ against static references, query-level routers, and CR$^2$ variants. Static baselines include \textbf{Always-1.7B}, \textbf{Always-4B}, \textbf{Always-8B}, and \textbf{Always-14B}. Query-level routers include \textbf{KNN} and \textbf{MLP} (RouterBench~\cite{hu2024routerbench}), \textbf{EmbedLLM}~\cite{zhuang2025embedllm}, and \textbf{LLMRank}~\cite{agrawal2025llmrankunderstandingllmstrengths}. These methods directly map queries to models, without the deployment-aligned factorization, i.e., separating local acceptance from post-deferral model selection.

We evaluate four main metrics. (i) \emph{routing accuracy} is the
fraction of correctly answered test queries. (ii) \emph{normalized
deployment cost} $\bar c$ is the average deployment cost relative
to Qwen3-$14$B. (iii) \emph{marginal false-acceptance risk} is the fraction of test
queries locally accepted despite a strictly better edge alternative under the teacher-estimated utility.
(iv) \emph{local acceptance rate} is the fraction of test queries accepted by the margin gate under a given $(\lambda,\alpha)$ operating point.

All routing methods using frozen query embeddings share the \texttt{all-MiniLM-L6-v2} encoder with mean pooling ($d=384$). The CR$^2$ margin gate is a two-layer MLP (hidden width $256$, dropout $0.1$) with a learnable temperature initialized at $0.1$. Training uses AdamW (learning rate $3\times10^{-4}$, batch size $256$, weight decay $0.01$, gradient clip $1.0$, Huber $\beta=0.1$) for $20$ epochs with $J$ log-uniform cost weights $\lambda^{(j)}\in[0.1,20]$ per step. Calibration uses $\alpha=0.010$ unless otherwise stated.

% ---------------------------------------------------------------------------
\begin{figure}[!t]
  \centering
  \includegraphics[width=0.48\textwidth]{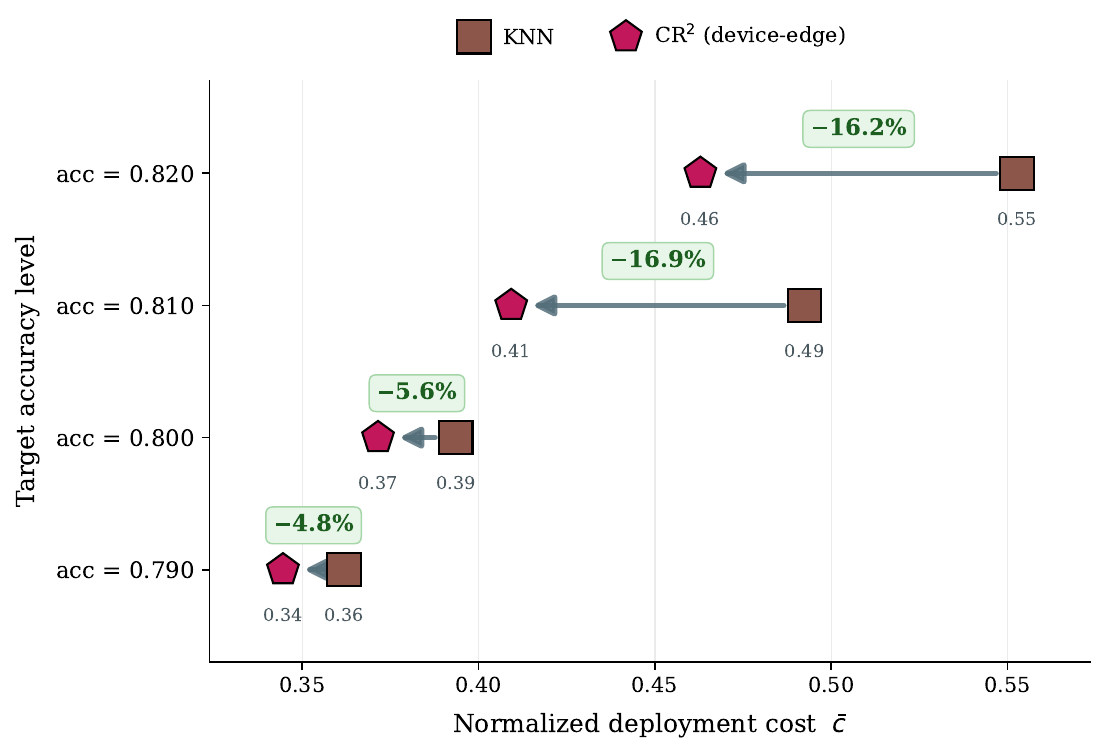}
  \caption{Fixed-accuracy cost comparison.}
  \label{fig:dumbbell_cost_saving}
\end{figure}

\begin{figure}[!t]
  \centering
  \includegraphics[width=0.48\textwidth]{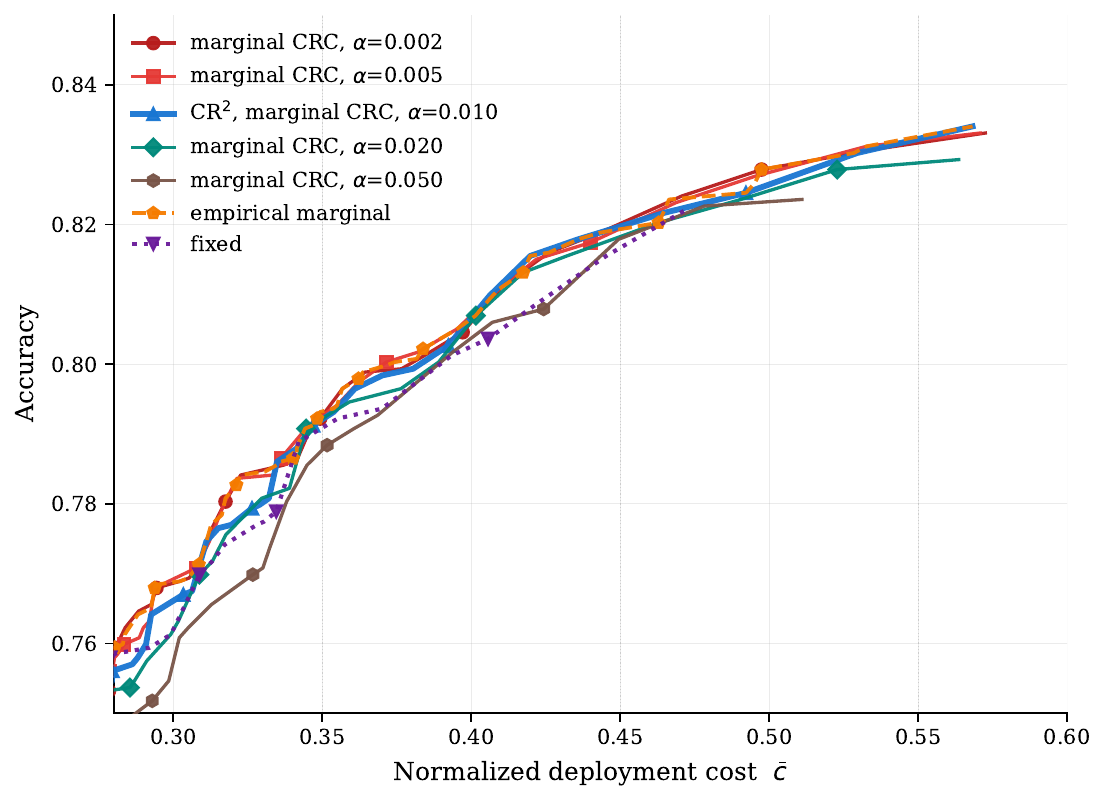}
  \caption{Accuracy--cost curves for threshold-selection rules.}
  \label{fig:calibration_methods}
\end{figure}

\subsection{Main Accuracy--Cost Pareto}
\label{sec:exp_main}

We compare CR$^2$ with baselines in the accuracy-cost plane. The x-axis in all Pareto figures is the sample-mean normalized deployment cost defined in Sec.~\ref{sec:sysmodel}. Router overhead is excluded and reported separately in Table~\ref{tab:router_complexity}. For Pareto envelopes, CR$^2$ pools CRC calibration levels $\alpha\in\{0.002,0.005,0.010,0.020,0.050\}$ over the sweep of $\lambda$; single-configuration results use $\alpha=0.010$ unless otherwise stated.

Fig.~\ref{fig:pareto_main} shows the main accuracy-cost comparison on the routing test split. In the zoomed range, CR$^2$ forms the strongest frontier among deployable methods, with \textbf{KNN}, \textbf{LLMRank}, and \textbf{EmbedLLM} as main competitors. CR$^2$ separates local acceptance from post-deferral edge model selection, aligning better with the asymmetric information structure of mobile-edge deployment and yielding a better accuracy–cost trade-off in the operating region.

Fig.~\ref{fig:dumbbell_cost_saving} presents a fixed-accuracy view, reporting minimum cost for each target accuracy. At $0.79$ and $0.80$ accuracy, CR$^2$ reduces
normalized cost vs. KNN by $4.8\%$ and
$5.6\%$; at $0.81$ and
$0.82$, the reductions increase to $16.9\%$
and $16.2\%$.

Table~\ref{tab:per_benchmark_by_cost} breaks results down by benchmark family at three cost targets. CR$^2$ achieves the best pooled accuracy in all cases. MLP is omitted at $\bar c=0.45$ and $\bar c=0.55$ due to limited operating range. Gains are not uniform across tasks: at $\bar c=0.35$, MLP is strongest on GPQA, while  CR$^2$ dominates GPQA at higher costs. Overall improvements thus reflect a stronger operating frontier rather than per-benchmark dominance.

\begin{figure}[!t]
  \centering
  \includegraphics[width=0.48\textwidth]{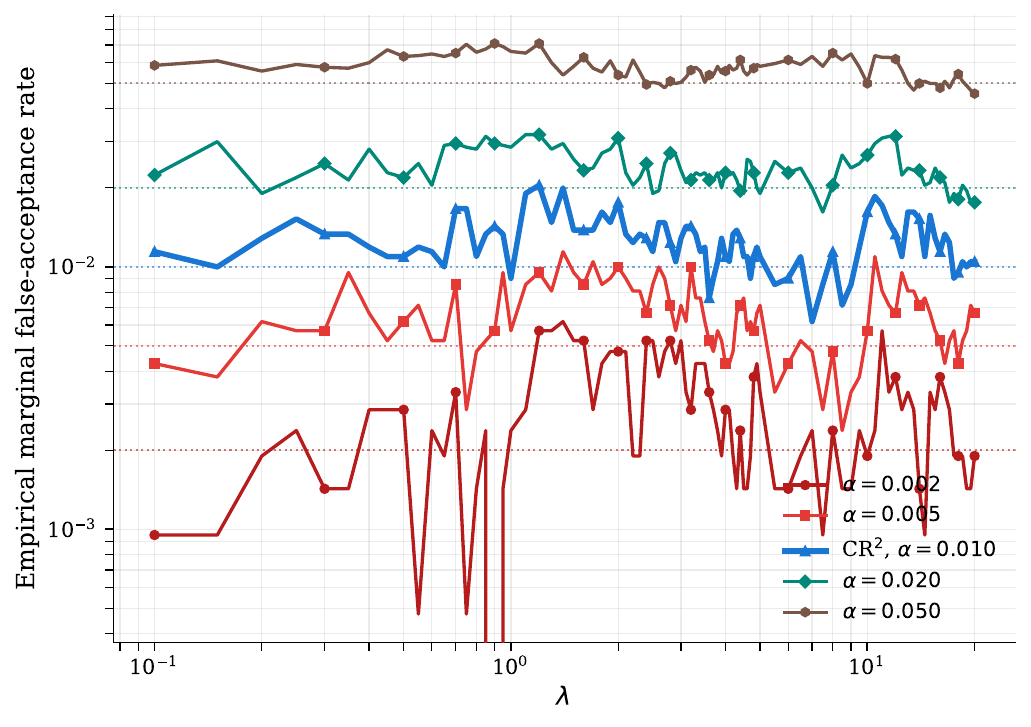}
  \caption{Marginal false-acceptance rate under CRC-calibrated thresholds.}
  \label{fig:risk_control}
\end{figure}

\begin{figure}[!t]
  \centering
  \includegraphics[width=0.48\textwidth]{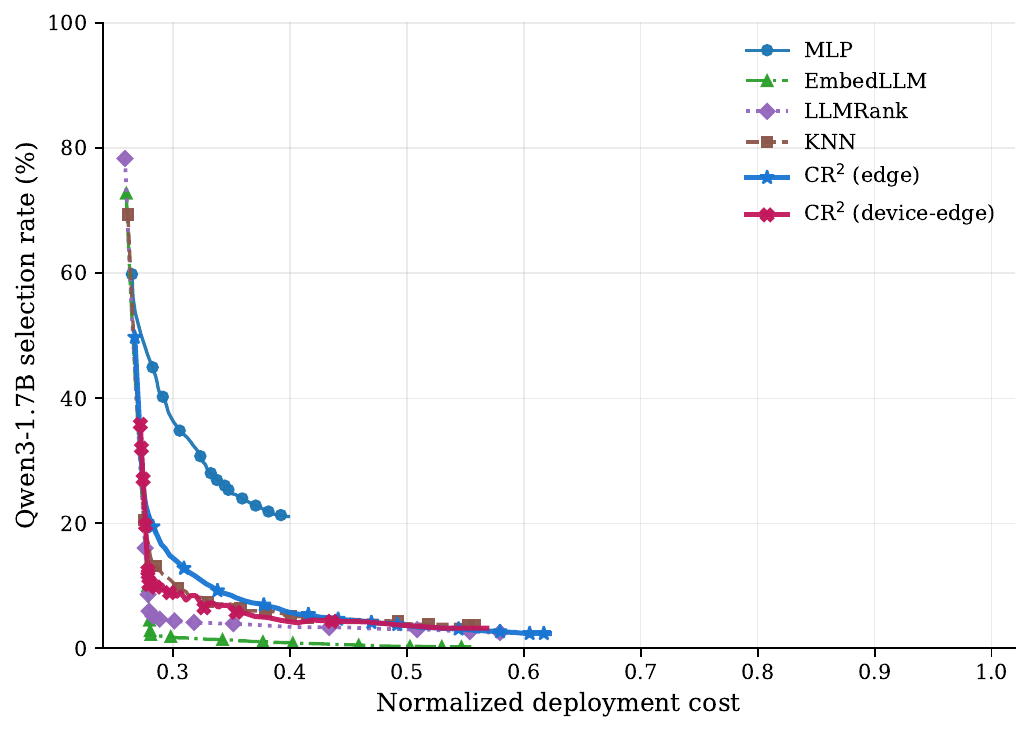}
  \caption{Local-model selection rate.}
  \label{fig:local_selection}
\end{figure}

\begin{table}[!t]
\centering
\caption{Per-benchmark accuracy at representative cost targets.
Blocks use nearest reachable operating point; ``--'' denotes an
unreachable target. Avg.\ is over all test queries.}
\label{tab:per_benchmark_by_cost}
\resizebox{0.48\textwidth}{!}{%
% Generated by scripts/table_tmc_per_benchmark.py
\begin{tabular}{lrrrrr}
\toprule
Method & MMLU & BBH & GPQA & MBPP & Avg \\
\midrule
\multicolumn{6}{l}{\emph{$\bar c = 0.35$}} \\
MLP & 0.762 & 0.850 & \textbf{0.624} & 0.718 & 0.738 \\
KNN & 0.806 & \textbf{0.872} & 0.518 & \textbf{0.821} & 0.754 \\
EmbedLLM & 0.799 & 0.857 & 0.533 & \textbf{0.821} & 0.752 \\
LLMRank & 0.802 & 0.868 & 0.515 & 0.769 & 0.738 \\
\textbf{CR$^2$ (device-edge)} & \textbf{0.812} & 0.864 & 0.544 & \textbf{0.821} & \textbf{0.760} \\
\midrule
\multicolumn{6}{l}{\emph{$\bar c = 0.45$}} \\
MLP & -- & -- & -- & -- & -- \\
KNN & 0.832 & \textbf{0.882} & 0.551 & 0.795 & 0.765 \\
EmbedLLM & 0.833 & 0.875 & 0.536 & \textbf{0.846} & 0.773 \\
LLMRank & 0.829 & 0.872 & 0.536 & 0.821 & 0.765 \\
\textbf{CR$^2$ (device-edge)} & \textbf{0.842} & 0.873 & \textbf{0.599} & 0.795 & \textbf{0.777} \\
\midrule
\multicolumn{6}{l}{\emph{$\bar c = 0.55$}} \\
MLP & -- & -- & -- & -- & -- \\
KNN & 0.849 & \textbf{0.882} & 0.562 & \textbf{0.897} & 0.798 \\
EmbedLLM & 0.851 & 0.877 & 0.522 & 0.846 & 0.774 \\
LLMRank & 0.848 & 0.873 & 0.547 & \textbf{0.897} & 0.792 \\
\textbf{CR$^2$ (device-edge)} & \textbf{0.858} & \textbf{0.882} & \textbf{0.628} & 0.846 & \textbf{0.804} \\
\bottomrule
\end{tabular}
}
\end{table}

\subsection{Calibration Under Risk Control}
\label{sec:exp_crc}
We next analyze the deployed gate under the calibrated risk criterion. Fixing the learned gate, we compare threshold-selection rules, vary the CRC calibration level $\alpha$, and report empirical marginal false-acceptance risk and local acceptance rate. Fig.~\ref{fig:calibration_methods} compares CRC calibration with alternative operating rules in the accuracy–cost plane. CRC remains competitive while explicitly controlling a false-acceptance target, rather than only tracing a favorable Pareto curve.

Fig.~\ref{fig:risk_control} evaluates risk control behavior. Empirical marginal false-acceptance risk preserves the expected ordering across $\alpha$ and stays within the corresponding range. Since curves are empirical estimates over finite test samples and a sweep of $\lambda$, they should not be interpreted as simultaneous deterministic bounds over the entire $\lambda$ sweep. Among tested values, $\alpha=0.010$ provides a good balance between conservativeness and frontier quality and is used as default; the full Pareto envelope reports the full $\alpha$ grid.

Fig.~\ref{fig:local_selection} shows local selection rates. One-shot routers exhibit sharp changes with cost weight, while CR$^2$ maintains a calibrated acceptance mechanism tied to the local–edge margin. Its local acceptance profile is close to the full-information CR$^2$ reference across most operating regimes, suggesting similar aggregate local selection behavior.

\subsection{Component Ablation}
\label{sec:exp_ablation}

Table~\ref{tab:ablation} summarizes component ablations
at a fixed cost anchor. Four observations emerge. First, adding the margin loss yields
the best accuracy ($0.8205$), indicating that
preserving the continuous margin geometry is beneficial beyond
binary sign prediction. Second, varying $w_{\rm mono}$ changes
single-anchor accuracy by at most $0.0006$, so the default is selected for the full operating curve rather than
this anchor. Third, increasing CRC calibration from $\alpha=0.002$ to $\alpha=0.050$ raises local
acceptance rate from $2.3\%$ to $9.4\%$ while reducing accuracy
from $0.8201$ to $0.8180$, and $\alpha=0.010$ is used as the
default. Fourth, deferred-branch results show that the
inclusive candidate set achieves the highest single-anchor accuracy,
while the edge-only setting is used in deployment.

\subsection{Deferred-Branch Comparison and Gate-Error Decomposition}
\label{sec:exp_deferred}

\begin{figure}[!t]
  \centering
  \includegraphics[width=0.48\textwidth]{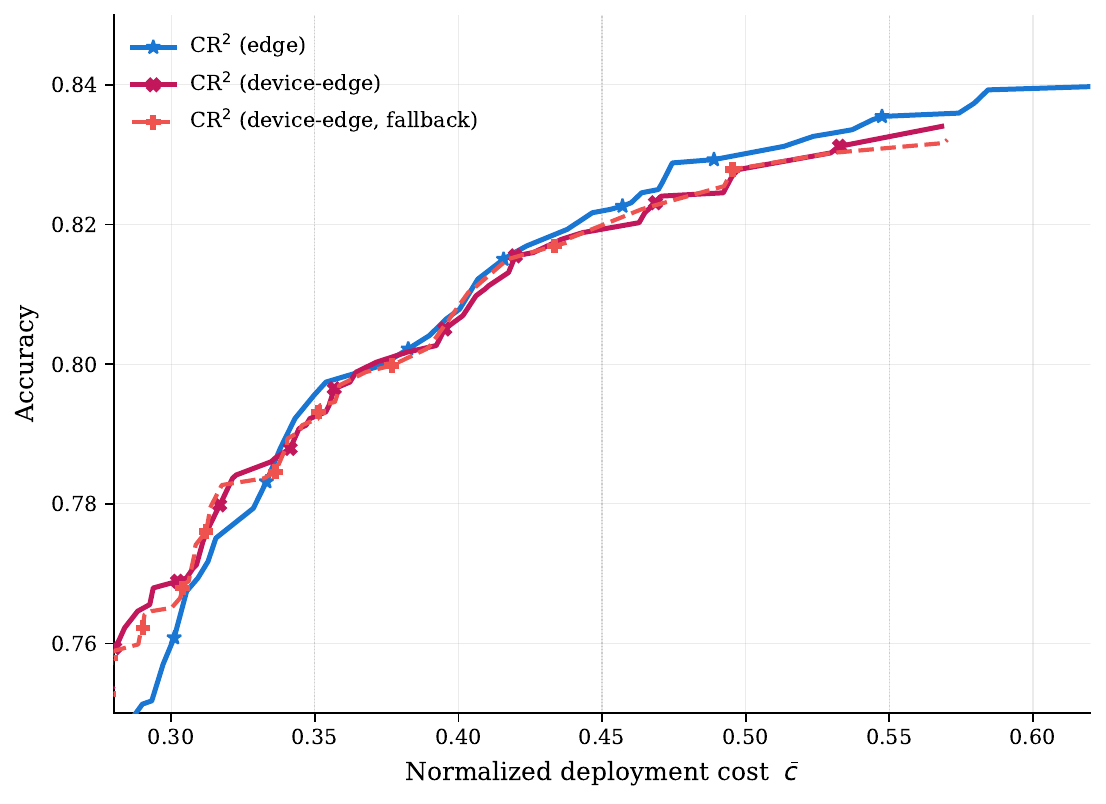}
  \caption{Deferred-branch variants.}
  \label{fig:ours_comparison}
\end{figure}

\begin{table}[!t]
\centering
\caption{Ablation at $\bar c=0.45$. Local rate is the fraction
of queries accepted by the margin gate.}
\label{tab:ablation}
% Generated by scripts/table_tmc_ablation.py
\begin{tabular}{lrr}
\toprule
Configuration & Accuracy & Local rate \\
\midrule
\multicolumn{3}{l}{\emph{Loss composition}} \\
sign loss only & 0.8202 & 2.3\% \\
+ margin loss & \textbf{0.8205} & 2.8\% \\
\textbf{+ margin + monotone (CR$^2$)} & 0.8199 & 2.5\% \\
\midrule
\multicolumn{3}{l}{\emph{$w_{\rm mono}$ sweep (FiLM, $J=8$)}} \\
$w_{\rm mono}=0$ & \textbf{0.8205} & 2.8\% \\
$w_{\rm mono}=0.5$ & 0.8201 & 2.5\% \\
\textbf{$w_{\rm mono}=1.0$ (CR$^2$)} & 0.8199 & 2.5\% \\
$w_{\rm mono}=2.0$ & 0.8201 & 2.4\% \\
\midrule
\multicolumn{3}{l}{\emph{Calibration method}} \\
CRC, $\alpha=0.002$ & \textbf{0.8201} & 2.3\% \\
\textbf{CRC, $\alpha=0.010$ (CR$^2$)} & 0.8197 & 4.3\% \\
CRC, $\alpha=0.050$ & 0.8180 & 9.4\% \\
empirical (marginal), $\alpha=0.010$ & 0.8198 & 4.3\% \\
fixed local rate, $f=0.10$ & 0.8167 & 10.6\% \\
fixed local rate, $f=0.20$ & 0.7970 & 21.9\% \\
\midrule
\multicolumn{3}{l}{\emph{Defer-path semantics}} \\
inclusive $\mathcal{M}_{\rm def}=\mathcal{M}$ & \textbf{0.8215} & 4.8\% \\
\textbf{edge-only $\mathcal{M}_{\rm edge}$ (CR$^2$)} & 0.8199 & 2.5\% \\
fallback to Qwen3-$4$B & 0.8195 & 2.5\% \\
\bottomrule
\end{tabular}

\end{table}

Fig.~\ref{fig:ours_comparison} compares the full-information CR$^2$ (edge) reference with the deployable CR$^2$ (device-edge) and  fallback variants under a fixed margin gate. The device-edge curve closely tracks the full-information reference across most cost ranges, and the fallback shows similar behavior, indicating that the remaining gap mainly stems from the first-stage gate rather than the choice of post-deferral branch.

\begin{figure}[!t]
  \centering
  \includegraphics[width=0.48\textwidth]{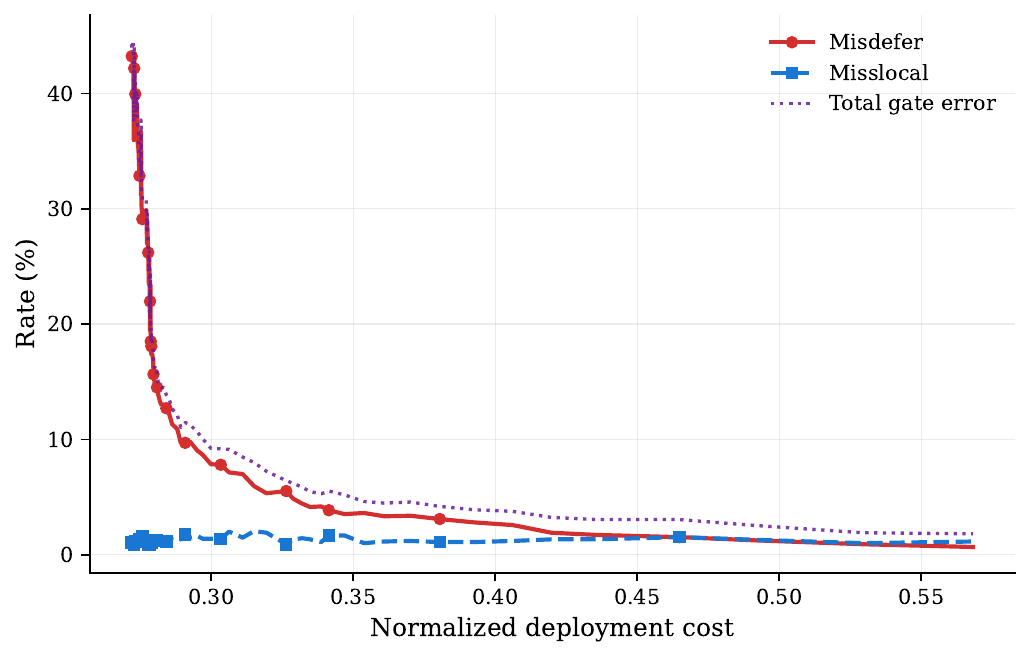}
  \caption{Gate-error decomposition.}
  \label{fig:misdefer}
\end{figure}

Fig.~\ref{fig:misdefer} further decomposes the first-stage gate errors into false deferrals and local acceptances. False local acceptance remains below $1.5\%$, while false deferral decreases from $3.56\%$ to $0.78\%$ as $\bar c$ increases from $0.35$  to $0.55$. Overall gate error drops from about $4.9\%$ to $1.9\%$, showing improved alignment with the full-information teacher at higher costs. The residual gap is thus mainly driven by over-escalation at lower costs, while false local acceptance remains limited.

\subsection{Router Complexity Measurement}
\label{sec:router_complexity}

Table~\ref{tab:router_complexity} summarizes the inference
complexity of the router heads, excluding the shared frozen
\texttt{all-MiniLM-L6-v2} encoder. This isolates the
per-query decision overhead incurred after the query
embedding has been computed. For CR$^2$ (device-edge), the
reported Params and FLOPs correspond to the first-stage margin gate only.

The non-parametric KNN router has the highest arithmetic cost. Although it has no trainable parameters, its per-query cosine search over roughly $18$k training embeddings requires $13.8$M FLOPs, far exceeding the fixed-size forward passes of all parametric routers. Among the parametric methods, LLMRank has the largest router head due to its feature, text, and fusion branches. The deployable CR$^2$ device-edge gate uses about $3.8\times$ fewer parameters and FLOPs than the full-information CR$^2$ (edge) reference, replacing four per-model binary heads with a single FiLM-conditioned margin gate. For deferred queries, adding the CR$^2$ (edge) selector gives about $996.6$k FLOPs including the gate, comparable to LLMRank and well below KNN; this path covers about $94$--$97\%$ of queries at the representative cost targets.

\begin{table}[!t]
\centering
\caption{Router-head parameters and FLOPs. The shared encoder is
excluded; CR$^2$ (device-edge) counts only the first-stage
margin gate.}
\label{tab:router_complexity}
% Generated by scripts/table_tmc_router_complexity.py
\begin{tabular}{lrr}
\toprule
Router & Params & FLOPs \\
\midrule
MLP & 235,204 & 469.2 k \\
KNN & --- & 13.826 M \\
EmbedLLM & 90,714 & 717.4 k \\
LLMRank & 608,516 & 1.215 M \\
\textbf{CR$^2$ (edge)} & 398,340 & 789.5 k \\
\textbf{CR$^2$ (device-edge)} & 104,706 & 207.1 k \\
\bottomrule
\end{tabular}

\end{table}

\section{Conclusion}

We presented CR$^2$, a two-stage collaborative routing framework for mobile-edge LLM inference that decouples a lightweight on-device margin gate from a state-aware edge utility selector. The proposed CRC-based calibration provides an explicit control for the marginal false-acceptance risk of local execution, while the edge selector preserves cost-aware, state-dependent model selection after deferral. Experiments on the routing task with profiled deployment costs show that the deployable device-edge realization closely tracks the full-information reference policy and reduces normalized deployment cost by up to $16.9\%$ at matched accuracy compared with the strongest evaluated baseline.

%=============================================================================

{\sloppy
\bibliographystyle{IEEEtran}
\bibliography{reference}
}

\end{document}